\documentclass[%
reprint,
superscriptaddress,
frontmatterverbose,
showpacs,preprintnumbers,
 amsmath,amssymb,
 aps,
pra,
]{revtex4-1}
\usepackage{subfigure}
\usepackage{xcolor}
\usepackage{graphicx}% Include figure files
\usepackage{dcolumn}% Align table columns on decimal point
\usepackage{bm}% bold math

\usepackage{float}

\begin{document}

%\preprint{APS/123-QED}

\title{Quantum Light on Demand}% Force line breaks with \\
\author{Mithilesh K. Parit}

\email{mithilesh.parit@gmail.com, mkp13ms113@iiserkol.ac.in}
\affiliation{Department of Physical Sciences, Indian Institute of Science Education and Research Kolkata, Mohanpur-741246, West Bengal, India}
\author{Shaik Ahmed}
\affiliation{Department of Humanities and Science, MLR Institute of Technology, Dundigal, Hyderabad-500043, Telangana, India}

\author{Sourabh Singh}
\affiliation{Department of Physical Sciences, Indian Institute of Science Education and Research Kolkata, Mohanpur-741246, West Bengal, India}

\author{P. Anantha Lakshmi}
%\altaffiliation{pochincherla03@gmail.com,  palsp@uohyd.ernet.in}
\email{pochincherla03@gmail.com,  palsp@uohyd.ernet.in}
\affiliation{School of Physics, University of Hyderabad, Hyderabad - 500046, Telangana, India}

\author{Prasanta K. Panigrahi}%
\email{pprasanta@iiserkol.ac.in, panigrahi.iiser@gmail.com}
\affiliation{Department of Physical Sciences, Indian Institute of Science Education and Research Kolkata, Mohanpur-741246, West Bengal, India}

\date{\today}% It is always \today, today,

\begin{abstract}
We demonstrate that light quanta of well defined characteristics can be generated in a coupled two-level system of three atoms. The quantum nature of light is controlled by the entanglement structure, discord, and monogamy of the system which leads to sub and superradiant behavior as well as sub-Poissonian statistics, at lower temperatures. Two distinct phases with different entanglement characteristics are observed with uniform radiation in one case and the other displaying highly focused and anisotropic radiation in far field regime. At higher temperatures, sub and superradiant light is found to persist in the absence of entanglement but with non-zero quantum discord, showing bunching of photons. It is shown that the radiation intensity can be a precise estimator of the inter-atomic distance of coupled two-level atomic systems. Our investigation shows for the first time, the three body correlation in the form of `monogamy score' controlling sub and superradiant nature of radiation intensity. 

\begin{description}

\item[keywords]%Use showkeys class option if keyword
 Two-level atoms, Supperradiance, Quantum discord, Entanglement

\end{description}
\end{abstract}
\maketitle

%%%%%%%%%%%%%%%%%%%%%%%%%%  body  %%%%%%%%%%%%%%%%%%%%%%%%%%
\section{Introduction}
Light of desired nature is much in demand for both fundamental \cite{aa1,gsa} and technological applications \cite{dwlg}. Study of coherent and incoherent sources of light, as well as sources generating single \cite{SPSs1, SPSs2, SPSs3} and entangled photons are subjects of intense investigation \cite{dep1,dep2,dep3,dep4,dep5,dep6,dep7,dep8,dep9,dep10}. The detection, characterization \cite{dep7,dep8}, and control of light have attracted significant attention. In this regard, light emission from entangled sources is being studied with particular interest to unveil the role of nonlocal quantum correlations on spontaneous emission, as also its superradiant character \cite{r17}, originally predicted by Dicke in 1954 \cite{r1}. Apart from dramatic enhancement of intensity, the emitted radiation provides much room for controlling its property and can provide a precise estimator of inter-atomic distance of coupled two-level atomic systems. %\newline

Dicke superradiance is the coherent spontaneous emission from a many-body system, owing its origin to the co-operative simultaneous interaction among its constituents, all of them experiencing a common radiation field \cite{r1}. The collective behavior of the ensemble arises from the coherent superposition and entanglement structure of the many-body wave function. The correlated structure can also show subradiant behavior due to destructive interference of the superposition states. It is interesting to note that some of the excited and ground states in the original study of Dicke are highly entangled \cite{r17}. Superradiance has been extensively studied in the literature, with the identification of a phase transition, separating the coherent phase of radiation from its  incoherent counterpart \cite{r2,r3,r4}.  It has attracted significant interest due to its possible applications, ranging from generation of X-ray lasers with high powers \cite{r5}, short pulse generation \cite{r6} to self-phasing in a system of classical oscillators \cite{r7}, to name a few. Super- and sub-radiant behavior  has been investigated experimentally in many physical systems \cite{r9,r10,cs, cs1, cs2, cs3}. In particular, Dimitrova et al. \cite{cs1} observed superradiance in a Bose-Einstein condensate (BEC). In yet another study, superfluidity of BEC along the axis of ring cavity has been shown to yield superradiant scattered photons \cite{cs3}. %\newline 

In the context of quantum information, it is of particular interest to explore how the behavior of radiation field gets affected for a collection of atoms, when the quantum states are correlated in different ways. This  allows  for optical probing  of quantum correlations (QCs) and aids in  quantifying QCs that may be  present in the system. It is well understood that depending on the nature of interactions of the multi-particle system, one can realize different types of entangled states \cite{gh, gh1} leading to different radiation characteristics. These atomic entangled states can find potential applications in quantum information processing \cite{dp}, for generating different entangled quantum states of light for quantum memories \cite{bc,rr}, quantum communication \cite{mde}, and quantum cryptography \cite{aks,ak},  among others. %\newline
    
  Two particle entanglement has been well characterized both for pure and mixed states \cite{VVd}, using different measures viz., von Neumann entropy and concurrence. Recently, concurrence \cite{chiru_sir2,chiru_sir} and quantum discord \cite{cm1} have been used for quantitatively characterizing entanglement governing the quantum phase transition occurring in an antiferromagnetic spin chain, consisting of weakly coupled dimers \cite{VVd,chiru_sir2,chiru_sir,cm1}. In comparison, the three particle entanglement is much less understood. It is known to exhibit stronger QCs as compared to the Bell states and  also shows stronger non-locality \cite{r12}. The entanglement structure of multiparticle states of different type are yet to be completely understood \cite{V_bhaskar1}. Here, we investigate the effect of entanglement, quantum discord, and monogamy relations on sub and superradiance of three two-level atom system. %\newline
  
    Recently, Wiegner et al. \cite{r17} have investigated the sub and superradiant characteristics of an N-atom system in a generalized W-state of the  form $\dfrac{1}{\sqrt{n}}|j,~ n-j> $, with $j$ atoms in the excited state and $(n - j)$ atoms  in the ground state, where the role of entanglement has been highlighted for pure states. In another study, the effect of quantum discord on sub and superradiant intensities in a system of X-type quantum states has  been investigated \cite{qd_int}, without taking into account the effect of finite temperatures. In the present study, we carry out a systematic investigation of the sub and superradiant properties of three dipole-coupled two level atoms and explore the effect  of transition frequency and coupling, on the resulting radiation pattern. The behavior of radiation field pattern as a function of concurrence and quantum discord is probed for gaining a physical understanding of the effect of different QCs on the emitted light. The role of QCs in producing highly collimated light, as well as completely uniform illumination is illustrated. The connection of entanglement on far field radiation pattern is demonstrated for line configuration. It is found that the other topologically distinct arrangement, the loop configuration, is not as effective as the line configuration for generating sub and superradiant light. The intensity pattern can be used to determine the inter-atomic distance. Further, it is found that intensity increases with monogamy of entanglement. The photon-photon correlation, as a function of system parameters, is found to yield sub and super-Poissonian characteristics, which can be controlled. %\newline
 
The  paper  is  organized  as  follows.   In  Sec. II,  the  Hamiltonian  for  the  system  of  three  identical  two-level atoms interacting via dipole-dipole coupling is introduced, using pseudo spin variables. We present the characteristics of the intensity of the emitted radiation from the three two-level atoms arranged in line configuration at non-zero temperature in Sec. III. Finally, we conclude with a  summary of the results and direction for future research. %\newline

%%%%\\\\\\\\\\\\\\\\\\\\\\\\\\\\\\\\\\\\\\\\\\\\\\\\\\\\\\\\\\\\\\\\\\\\\\
%%%% THEORY AND MODEL
%%%%\\\\\\\\\\\\\\\\\\\\\\\\\\\\\\\\\\\\\\\\\\\\\\\\\\\\\\\\\\\\\\\\\\\\\\

\section{Model}
We consider a system of three coupled identical atoms, where the excited state $\left|e_i\right>$ and the ground state $\left|g_i\right>$, $i = 1,~ 2,~ 3$ are separated by an energy interval $\hbar \omega$. The Hamiltonian for the system of three identical two-level atoms coupled through dipole-dipole interaction is given by,

\begin{equation}\label{eq:Hamiltonian}
H=\hbar {\sum^{3}_{i=1}}\omega_i S^{z}_{i}+ \hbar\sum^{3}_{i\neq j=1}\Omega_{ij}S^{+}_{i}S^{-}_{j}.
\end{equation}
The first term describes the unperturbed energy of the system and the second term represents the dipole-dipole interaction  between the ground state of one atom and the excited state of another atom, where,  $\Omega_{ij} $, the dipole-dipole interaction strength, which is a function of the inter-atomic separation `$ d $'.  The nature of dipole-dipole interaction prohibits interaction between two atoms which are both in excited/ground state. In the above, $S^{+}_{i} = (|1\rangle \langle 0|)_{i}$  and $S^{-}_{i}=(|0 \rangle \langle 1|)_{i}$  are the raising and lowering operators of the $i^{th}$ atom in the spin representation. Our system is closed and non-interactive with environment, which can be extended to open system dynamics \cite{SB1, SB2}. %\newline

We investigate the intensity emitted by three atom system, in the far field zone i. e., $|\vec{r}|>>d$; where $d$ is spacing between the atoms and $\vec{r}$  denotes the position of the detector to record the photons emitted by the atoms in the far field regime. The exact analytical expression for intensity is presented in supplementary material. In the ensuing sections, we investigate the intensity pattern resulting from the line configurations as a function of the system parameters, as well as the observation angle and temperature.

%%%%\\\\\\\\\\\\\\\\\\\\\\\\\\\\\\\\\\\\\\\\\\\\\\\\\\\\\\\\\\\\\\\\\\\\\\
%%% THE INTENSITY CHARACTERISTICS OF THE LINE-CONFIGURATION
%%%%\\\\\\\\\\\\\\\\\\\\\\\\\\\\\\\\\\\\\\\\\\\\\\\\\\\\\\\\\\\\\\\\\\\\\\
\section{The intensity characteristics of the line-configuration} 
\begin{figure}[ht!]
\centering
\includegraphics[width=9.30 cm,height=7.50 cm]{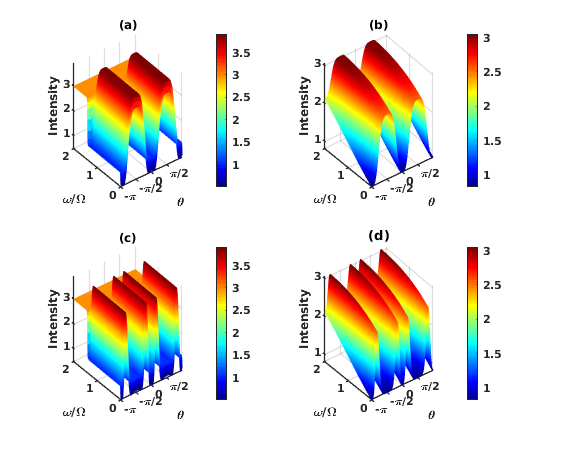} 
\caption{(Color online) The variation of radiation intensity as a function of $\frac{\omega}{\Omega}$ and observation angle is depicted for two different temperatures and ratio of emission wavelength and inter-atomic spacing ($\frac{\lambda}{d}$). Panels $a$ and $b$ show the intensity variation at $\frac{\lambda}{d}=2$ for (a) $k_BT=5\times10^{-3}\hbar\Omega$ and (b) $k_BT=\hbar\Omega$. Panels $c$ and $d$ show the radiation intensity at $\frac{\lambda}{d}=1$ for (c) $k_BT=5\times10^{-3}\hbar\Omega$ and (d) $k_BT=\hbar\Omega$, clearly revealing two distinct phases and interference effect.}
\label{fig:Line w&d_theta&w}
\end{figure}

\begin{figure}[H]
\centering \includegraphics[width=9.0cm,height=6.0cm]{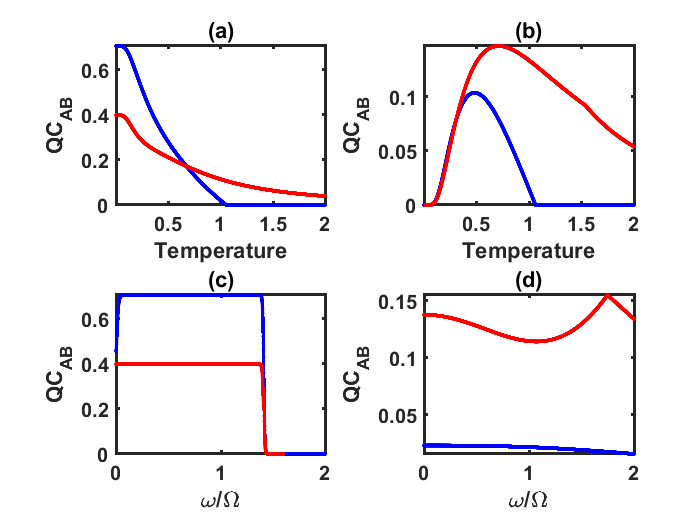}
\caption{(Color online) Panels $a$ and $b$ show the variation of QCs (Concurrence (blue) and Discord (red)) as a function of temperature ($\times \frac{\hbar\Omega}{k_B}$) for (a) $\frac{\omega}{\Omega}=1$ and (b) $\frac{\omega}{\Omega}=2$. Panels $c$ and $d$ show the variation of QCs as function of $\frac{\omega}{\Omega}$ for (c) $k_BT=5\times10^{-3}\hbar\Omega$ and (d) $k_BT=\hbar\Omega$, showing vanishing of QCs at high $\frac{\omega}{\Omega}$ and high temperatures.
}
\label{fig:Line QC}
\end{figure}

In an earlier work, the role of entanglement on super and subradiant behavior for the three atom system,  with a zero net dipole moment, was studied \cite{r17}.  Here, we have generalized this study, exhibiting the presence of QCs and their physical effect for the three atom system. Fig.\ref{fig:Line w&d_theta&w} depicts the periodic variation in the intensity from super to subradiant behavior as a function of ratio of transition frequency and dipole coupling ($\frac{\omega}{\Omega}$) and observation angle for two temperatures, and ratio of emission wavelength and inter-atomic spacing ($\frac{\lambda}{d}$). This reflects the subtle interference effects present in the three particle system. For high $\frac{\omega}{\Omega}$ and at low temperatures, a phase with uniform light emission is seen, separated from a non-uniform intensity with periodic modulations. The uniform phase (emission) of radiation arises when both entanglement and discord vanish, as is evident from Figs. \ref{fig:Line w&d_theta&w}(a), \ref{fig:Line w&d_theta&w}(c), and \ref{fig:Line QC}(c). A smooth crossover connects the two phases. The plot of crossover of eigen-energies is presented in the supplementary material (Fig. S1 (b)). The uniform phases vanish at higher temperatures as seen in Figs. \ref{fig:Line w&d_theta&w}(b) and \ref{fig:Line w&d_theta&w}(d). %\newline

%%%%||||||||||||||||||||||||||||||||||||||||||||||||||||||||||||||||||||||||
%%%%||||||||||||||||||||||||||||||||||||||||||||||||||||||||||||||||||||||||

%%%%||||||||||||||||||||||||||||||||||||||||||||||||||||||||||||||||||||||||
%\begin{verbatim}
%\end{verbatim}
%%%%||||||||||||||||||||||||||||||||||||||||||||||||||||||||||||||||||||||||

\begin{figure}[H]
\centering
\includegraphics[width=9.50 cm,height=4.70 cm]{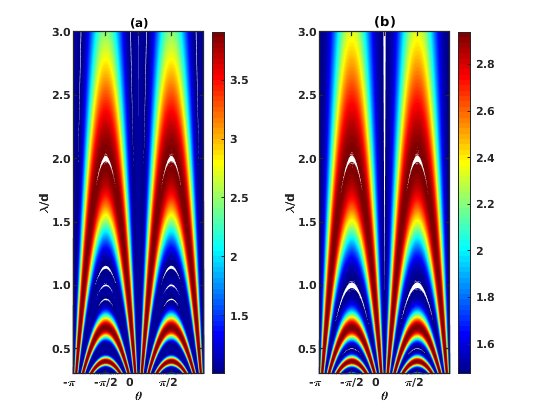} 
\caption{(Color online) The variation of intensity is shown as a function of observation angle and $\frac{\lambda}{d}$ at fixed $\frac{\omega}{\Omega}~(=1)$ for (a) $k_BT=5\times10^{-3}\hbar\Omega$ and (b) $k_BT=\hbar\Omega$, clearly showing the interference effect.}
\label{fig:Line theta d}
\end{figure}

To understand the intensity profile, for the given system, it is imperative to know the variation of QCs with temperature and transition frequency. In Fig. \ref{fig:Line QC}, panels $a$ and $b$ show the behavior of concurrence \cite{ref:conc} and quantum discord \cite{ref:QD, ref:H&V, ref:QD Luo} as a function of temperature, for two different values of the transition frequencies, while panels $c$ and $d$ show the variation of QCs as function of transition frequency for two different temperatures. For small value of transition frequency (temperature), increasing the temperature (transition frequency) leads to reduction in the value of both concurrence and discord, with concurrence vanishing for $k_BT=\hbar\Omega$ but discord remaining non-zero beyond this temperature. Thus, the intensity pattern at small $k_BT$ and small values of $\omega$ observed in Fig. \ref{fig:Line w&d_theta&w} is predominantly due to the high amount of QCs present in the system. This result also confirms that even for $k_BT>\hbar\Omega$, the superradiant behavior is present, albeit with reduced intensity in the absence of concurrence but with non-zero discord as is evident from Figs. \ref{fig:Line w&d_theta&w} and \ref{fig:Line QC}. This explicates the physical significance of quantum discord. The blue region in the plot represents subradiant behavior, while the red regions correspond to superradiant behavior. Fig. \ref{fig:Line w&d_theta&w} clearly shows that the wavelength of the emitted radiation and observation angle play significant role in finding superradiant light in the far-field domain. The intensity is maximum at observation angle $\theta=\pm \frac{\pi}{2}$ only in the vicinity of $\frac{\lambda}{d}=\frac{2}{5}$, $\frac{\lambda}{d}=\frac{2}{3}$, and $\frac{\lambda}{d}=2$. The intensity pattern for different combinations of $\theta$ and $\frac{\lambda}{d}$ is shown in Fig. \ref{fig:Line theta d}.

%%%%||||||||||||||||||||||||||||||||||||||||||||||||||||||||||||||||||||||||
%%%%||||||||||||||||||||||||||||||||||||||||||||||||||||||||||||||||||||||||

The behavior of intensity as a function of observation angle and $\frac{\lambda}{d}$ at fixed $\frac{\omega}{\Omega}~(=1)$ for different temperatures is shown in Fig. \ref{fig:Line theta d}.  Sub and superradiant nature of radiation is observed at all observation angles (except for $\theta=n\pi$). For $\theta=\frac{\pi}{2}$, the intensity observed is super-radiant in the vicinity of $\frac{\lambda}{d}=\frac{2}{5}$, $\frac{\lambda}{d}=\frac{2}{3}$, and $\frac{\lambda}{d}=2$ only. It can be clearly seen that the maximum value of intensity is higher for panel $a$ than that of  panel $b$. This can be attributed to higher QCs present for ``$\frac{\omega}{\Omega}=1$ and $k_BT=5\times10^{-3}\hbar\Omega$'' as compared to the QCs  for ``$\frac{\omega}{\Omega}=1$ and $k_BT=\hbar\Omega$'' (see Fig. \ref{fig:Line QC}). This result is significant in light of the fact that it provides a method to find the inter-atomic distance of equally spaced array of atoms. For example, for given emission wavelength $\lambda$ the superradiant intensity will be observed at specific angles. As $\lambda$ and $\theta$ are known and from Fig. \ref{fig:Line theta d}, the relation between $\frac{\lambda}{d}$ and $\theta$ can be used to estimate $d$. Therefore, on observing the emitted photons at different observation angles, one can infer about the inter-atomic distance of the system. The intensity pattern as a function of $\frac{\lambda}{d}$ and $\theta$ for $3$, $4$, and $5$ atoms is presented in supplementary information (Fig. S3). It is evident from Fig. \ref{fig:Line theta d} that the behavior of intensity remains same at higher temperature, albeit with reduced intensity. %\newline

\begin{figure}[ht!]
\centering
\includegraphics[width=9 cm,height=5.0 cm]{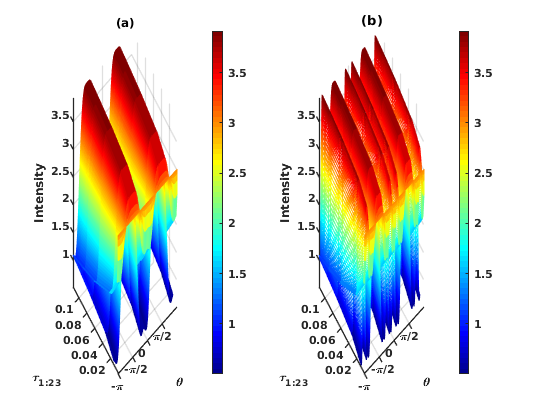} 
\caption{(Color online) Panels $a$ and $b$ show intensity variation with respect to monogamy score ($\tau_{1:23}$) of negativity at $k_BT=5\times10^{-3}\hbar\Omega$ for (a) $\frac{\lambda}{d}=2$ and (b) $\frac{\lambda}{d}=\frac{2}{3}$, showing the increase of intensity with increase in monogamy score.}
\label{fig:Line Neg tangle theta d}
\end{figure}

The variation of intensity as functions of monogamy score \cite{mngy1, mngy2} of negativity ($\tau_{1:23}$) \cite{ref:neg} is depicted in Fig. \ref{fig:Line Neg tangle theta d} at $k_BT=5\times10^{-3}\hbar\Omega$. It is observed that higher is the monogamy score of negativity, stronger is the superradiance. It represents shareability of entanglement (QCs) among entities, thus, as monogamy increases, the shareability of entanglement increases and thereby quantum coherence increases, explaining the superradiant intensity. This clearly shows the relevance of monogamy relations in a physical scenario. This may find application in quantum cryptography, as higher is the intensity observed from a system of array of atoms ($\geq3$), more secure it is.

%%%%||||||||||||||||||||||||||||||||||||||||||||||||||||||||||||||||||||||||
%%%%||||||||||||||||||||||||||||||||||||||||||||||||||||||||||||||||||||||||

%%%%||||||||||||||||||||||||||||||||||||||||||||||||||||||||||||||||||||||||
\begin{figure}[ht!]
\centering
\includegraphics[width=9.5 cm,height=7.0 cm]{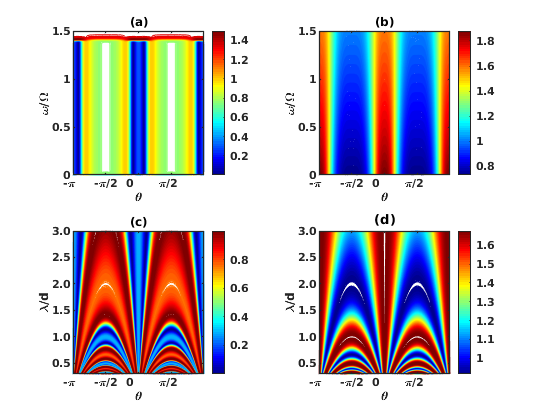} 
\caption{(Color online) Panels $a$ and $b$ show photon-photon correlation for $\frac{\lambda}{d}=2$ as function of $\frac{\omega}{\Omega}$ and observation angle for (a) $k_BT=5\times10^{-3}\hbar\Omega$ and (b) $k_BT=\hbar\Omega$. Panels $c$ and $d$ show photon-photon correlation for $\frac{\omega}{\Omega}=1$ as function of wavelength of emitted radiation and observation angle for (c) $k_BT=5\times10^{-3}\hbar\Omega$ and (d) $k_BT=\hbar\Omega$, clearly indicating the sub and super-Poissonian statistics.}
\label{fig:ph_ph corr}
\end{figure}
%%%%||||||||||||||||||||||||||||||||||||||||||||||||||||||||||||||||||||||||

    The behavior of photon-photon correlation ($g^2(0)$) as a function of observation angle and $\frac{\omega}{\Omega}$/$\frac{\lambda}{d}$ is depicted in Fig. \ref{fig:ph_ph corr}. It can be clearly seen that photon statistics at higher temperature ($k_BT=\hbar\Omega$) is mostly super-Poissonian. It is evident from Fig. \ref{fig:ph_ph corr}(c) that for ``$\frac{\omega}{\Omega}=1$ and $k_BT=5\times10^{-3}\hbar\Omega$'' intensity pattern follows sub-Poissonian behavior for all values of $\frac{\lambda}{d}$ and $\theta$. It is to be noted that photons emitted from entangled sources display quantum nature at lower temperatures while at higher temperatures classical behavior is expected and show bunching of photons, even if quantum discord does not vanish. The photon-photon correlation for maximally entangled (concurrence=1) sources is zero, as can be seen in supplementary material Fig. S4. The  analytical expression for the photon-photon correlation is presented in supplementary material. It is evident that photon-statistics can be controlled parametrically. %\newline

\section{Conclusion}
In conclusion, the emitted radiation from a coupled three particle system is shown to be a rich source of light of desired characteristics. Intensity pattern can infer about inter-atomic distance of equally spaced atomic systems by observing the radiation at different observation angles (Fig. \ref{fig:Line theta d}). It can be both uniform and highly focused in the far field regime in a controlled manner. The highly focused light owes its origin to QCs and can find application for lithography \cite{dwlg} and other technological applications. We have obtained the exact expression for the radiation intensity and photon-photon correlation in the far field domain for three atoms at finite temperature. The photon-photon correlation demonstrates the sub-Poissonian (anti-bunching) and super-Poissonian (bunching) statistics of emitted photons and the fact that it can be controlled by tuning the system parameters. The effect of quantum correlations on emitted light, viz., concurrence, quantum discord, and monogamy score is explicitly demonstrated. The radiative behavior shows dramatic variation as a function of concurrence, quantum discord, and monogamy score of negativity, revealing the role of distinct QCs, thereby providing an optical probe for studying the quantum characteristics of  emitting sources. Apart from revealing the physical signature of entanglement and quantum discord on the behavior of light, our investigation shows for the first time, the effect of three body correlation in the form of `monogamy score' on superradiance. The shareability of QCs in a multipartite system is recorded through monogamy, which directly affects the radiation intensity. The fact that monogamy physically represents `sharing' of quantum correlation in a multi-party channel and is found here to directly control the superradiant character of the intensity suggests the use of superradiance as a `QC sharing witness' in a multi-party network. This may find application in the use of  multiparty entanglement for secure information sharing and communication. The precise `witness' character of monogamy score and the measurable property of superradiance reflecting the same are currently under investigation and will be reported elsewhere. Conditions under which hyper-radiance can be achieved is under investigation and also extension of this study to open quantum systems. %\newline

\section{Acknowledgements}
Mithilesh K. Parit acknowledges discussion with  Dr. Chiranjib Mitra and Department of Science and Technology, New Delhi, India for providing the DST-INSPIRE fellowship during his stay at IISER Kolkata. %\newline

\pagebreak

\onecolumngrid

\section*{Supplementary Information: Quantum Light on Demand}% Force line breaks with \\
\author{{Mithilesh K. Parit$^{1, 2, *}$, Shaik Ahmed$^3$, Sourabh Singh$^1$, P. Anantha Lakshmi$^{4, \dagger}$, Prasanta K. Panigrahi$^1$}\\
%\skiplinehalf
{\small \em $^1$Department of Physical Sciences, Indian Institute of Science Education and Research Kolkata, Mohanpur-741246, West Bengal, India\\
$^2$Laboratoire Charles Fabry, Institut d'Optique, CNRS, Universite Paris Sud 11,
2 Avenue Augustin Fresnel, F-91127 Palaiseau Cedex, France \\
$^3$Department of Humanities and Science , MLR Institute of Technology, Dundigal, Hyderabad-500043, Telangana, India\\
$^4$School of Physics, University of Hyderabad, Hyderabad - 500046, Telangana, India
}}
%\begin{document} 
\maketitle 
\date{\today}

\subsection{The model}
We consider a system of three coupled identical atoms, where the excited state $\left|e_i\right>$ and the ground state $\left|g_i\right>$, $i = 1, ~2,~ 3$ are separated by an energy interval $\hbar \omega$. The Hamiltonian for the system of three identical two-level atoms coupled through dipole-dipole interaction is given by,

\begin{equation}\label{eq:Hamiltonian}
H={\hbar \sum^{3}_{i=1}}\omega_i S^{z}_{i}+\hbar \sum^{3}_{i\neq j=1}\Omega_{ij}S^{+}_{i}S^{-}_{j}.
\end{equation}

%The schematic representation of the two configurations are shown below:

%%%%|||||||||||||||||||||||||||||||||||||||||||||||||||||||||||||||||||||||||||
\begin{figure}[H]
\centering
\subfigure[]{
\includegraphics[width=8.0 cm,height=7.0 cm]{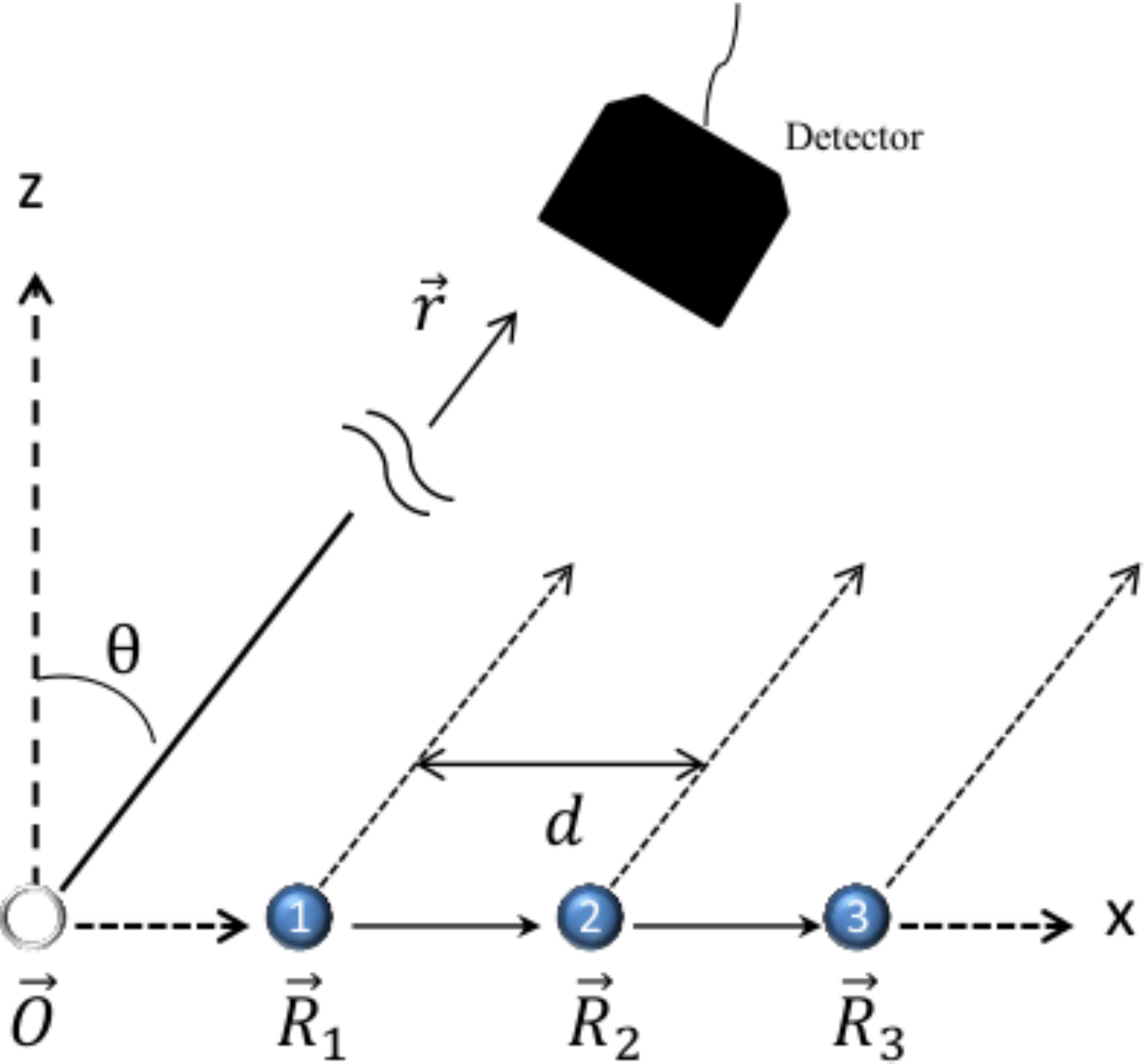}
\label{fig:one a}
}
\subfigure[]{
\includegraphics[width=8.0 cm,height=6.0 cm]{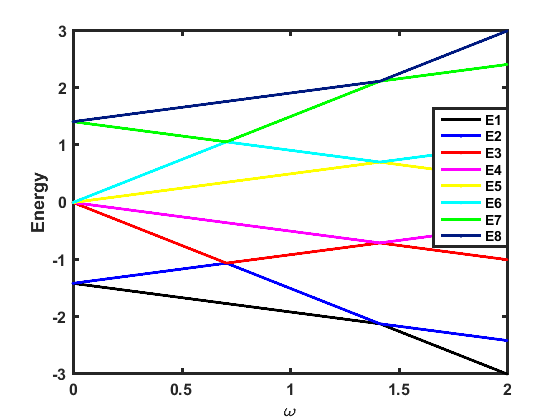}
\label{fig:one b}
}
\caption[]{(Color online) (a) Schematic diagram of the system in the line configurations with identical two-level atoms localized at positions $\bar{R}_{1}$ to $\bar{R}_{3}$. A detector is placed at position $ \bar{r} $ to record the photons emitted by the atoms in the far field regime.(b) Eigenenergies as a function of $\frac{\omega}{\Omega}$ for line configuration revealing cross over.}
\label{fig:One}
\end{figure}
%%%%|||||||||||||||||||||||||||||||||||||||||||||||||||||||||||||||||||||||||||

In the subsequent sections, we have calculated the eigenvalues of the Hamiltonian and their corresponding eigenstates for line configuration. The exact analytical expressions for intensity and photon-photon correlation of three atoms arranged along a line are derived.

\subsection[]{The analytical expression for intensity and photon-photon correlation in line configuration}

At thermal equilibrium, the quantum state of a three atom system is a weighted superposition of all the eigenstates. For simplicity, we consider the  transition frequencies of  all the three atoms to be the same, $\omega_{1}=\omega_{2}=\omega_{3}=\omega$ and the nearest neighbor dipole-dipole interactions $ \Omega_{12}=\Omega_{23}=\Omega$ and $\Omega_{13}=0$. By diagonalizing the Hamiltonian $H$, we can obtain all the eigenvalues $ \epsilon_{i} $ and their corresponding eigenstates $|\psi_{i} \rangle $.  The eigenvalues, $\epsilon_i$,  in the line configuration are
\begin{align}
 \epsilon_{1}&=\frac{-3\hbar \omega}{2};~\epsilon_{2}=-\sqrt{2}\hbar \Omega-\frac{\hbar \omega}{2};~\epsilon_{3}=-\frac{\hbar \omega}{2};~ \epsilon_{4}=\sqrt{2}\hbar \Omega-\frac{\hbar \omega}{2}\nonumber \\
\epsilon_{5}&=-\sqrt{2}\hbar \Omega+\frac{\hbar \omega}{2};~\epsilon_{6}=\frac{\hbar \omega}{2}; ~ \epsilon_{7}=\sqrt{2}\hbar \Omega+\frac{\hbar \omega}{2};~\epsilon_{8}=\frac{3\hbar \omega}{2}
\label{eval1}
\end{align}
and the corresponding  eigenstates, $|\psi_{i} \rangle $, of the system are given by,
\begin{align} 
|\psi_{1} \rangle &= |g_{1}g_{2}g_{3} \rangle;~~ |\psi_{2} \rangle = \frac{1}{2}\left[|e_{1}g_{2}g_{3} \rangle -\sqrt{2}|g_{1}e_{2}g_{3} \rangle +|g_{1}g_{2}e_{3}\rangle \right]\nonumber \\
 |\psi_{3} \rangle &= \frac{1}{\sqrt{2}}\Big[|g_{1}g_{2}e_{3} \rangle - |e_{1}g_{2}g_{3} \rangle \Big];~~ |\psi_{4} \rangle = \frac{1}{2}\left[|e_{1}g_{2}g_{3} \rangle +\sqrt{2}|g_{1}e_{2}g_{3}  \rangle +|g_{1}g_{2}e_{3} \rangle \right]\nonumber \\
 |\psi_{5} \rangle &= \frac{1}{2}\left[|e_{1}e_{2}g_{3} \rangle -\sqrt{2}|e_{1}g_{2}e_{3} \rangle +|g_{1}e_{2}e_{3}  \rangle \right];~~ |\psi_{6} \rangle = \frac{1}{\sqrt{2}}\Big[|g_{1}e_{2}e_{3} \rangle - |e_{1}e_{2}g_{3} \rangle \Big]  \nonumber \\
 |\psi_{7} \rangle &= \frac{1}{2}\left[|e_{1}e_{2}g_{3} \rangle +\sqrt{2}|e_{1}g_{2}e_{3} \rangle +|g_{1}e_{2}e_{3}  \rangle \right];~~ |\psi_{8}\rangle = |e_{1}e_{2}e_{3} \rangle. \nonumber \\
 \label{evec1}
 \end{align}
 The thermal density matrix of the system is given by
 \begin{equation}
\rho_{ABC}=\frac{\sum_{i=1}^{8}\left|\psi_i\right>\left<\psi_i\right|\text{exp}\left(-\beta \epsilon_i \right)}{\text{Tr}\left(\sum_{i=1}^{8}\left|\psi_i\right>\left<\psi_i\right|\text{exp}\left(-\beta \epsilon_i \right)\right)}.
\label{rho1}
\end{equation}
By combining Eqs. (\ref{eval1}) to (\ref{rho1}), one can obtain the thermal density matrix of the form,

\begin{equation}
  \rho_{ABC}(T)=\dfrac{1}{Z}\begin{bmatrix}
\rho_{11} & 0 & 0 & 0 & 0  & 0 & 0 & 0 & \\
0 & \rho_{22} & \rho_{23} & 0 & \rho_{25}  & 0  & 0 & 0  \\
0 & \rho_{32} & \rho_{33} & 0 & \rho_{35} & 0 & 0 & 0  \\
0 & 0 & 0 & \rho_{44} & 0 & \rho_{46} & \rho_{47} &  0  \\
0 & \rho_{52} & \rho_{53} & 0 & \rho_{55} & 0 & 0 & 0 \\
0 & 0 & 0 & \rho_{64} & 0 & \rho_{66} & \rho_{67}  & 0 \\
0 & 0 & 0 & \rho_{74} & 0 & \rho_{76} & \rho_{77} & 0\\
0 & 0 & 0  & 0 & 0 & 0 & 0 & \rho_{88} \\
\end{bmatrix} 
\label{line_density}
\end{equation}  
where the partition function $Z$ is given by

\begin{align}
    Z=2~ \text{cosh}\left(\frac{\hbar \omega }{2~ k_BT}\right) \left(1+8 ~\text{cosh}\left(\frac{\sqrt{2}\hbar \Omega   }{k_BT}\right)+2~ \text{cosh}\left(\frac{\hbar \omega }{k_BT}\right)\right).
\end{align}
The non-vanishing elements of density matrix $\rho_{ABC}(T)$ are given by,

\begin{align}
\rho_{11}=\text{exp}\left(-\frac{3 \hbar \omega }{2 k_BT}\right); ~~~
\rho_{22}=\text{exp}\left(-\frac{\hbar \omega }{2~ k_BT}\right) \left(1+2~ \text{cosh}\left(\frac{\sqrt{2} \Omega  \hbar }{k_BT}\right)\right); \nonumber \\[5pt]
\rho_{23}=-2 \sqrt{2}~ \text{exp}\left(-\frac{\hbar \omega }{2~ k_BT}\right) \text{sinh}\left(\frac{\sqrt{2} \Omega  \hbar }{k_BT}\right);~ \rho_{25}=\text{exp}\left(-\frac{\hbar \omega }{2 k_BT}\right) \left(-1+2~ \text{cosh}\left(\frac{\sqrt{2} \Omega  \hbar }{k_BT}\right)\right); \nonumber \\[5pt]
 \rho_{32}=-2 \sqrt{2}~ \text{exp}\left(-\frac{\hbar \omega }{2~ k_BT}\right) \text{sinh}\left(\frac{\sqrt{2} \Omega  \hbar }{k_BT}\right);~ \rho_{33}=4~ \text{exp}\left(-\frac{\hbar \omega }{2 k_BT}\right) \text{cosh}\left(\frac{\sqrt{2} \Omega  \hbar }{k_BT}\right); \nonumber \\[5pt]
 \rho_{35}=-2 \sqrt{2}~ \text{exp}\left(-\frac{\hbar \omega }{2~ k_BT}\right) \text{sinh}\left(\frac{\sqrt{2} \Omega  \hbar }{k_BT}\right);~ \rho_{44}=\text{exp}\left(\frac{\hbar \omega }{2 k_BT}\right) \left(1+2~ \text{cosh}\left(\frac{\sqrt{2} \Omega  \hbar }{k_BT}\right)\right); \nonumber
 \end{align}
 
 \begin{align}
 \rho_{46}=-2 \sqrt{2}~ \text{exp}\left(\frac{\hbar \omega }{2~ k_BT}\right) \text{sinh}\left(\frac{\sqrt{2} \Omega  \hbar }{k_BT}\right);~ \rho_{47}=\text{exp}\left(\frac{\hbar \omega }{2 k_BT}\right) \left(-1+2~ \text{cosh}\left(\frac{\sqrt{2} \Omega  \hbar }{k_BT}\right)\right); \nonumber \\[5pt]
 \rho_{52}=\text{exp}\left(-\frac{\hbar \omega }{2~ k_BT}\right) \left(-1+2 \text{cosh}\left(\frac{\sqrt{2} \Omega  \hbar }{k_BT}\right)\right);~ \rho_{53}=-2 \sqrt{2}~ \text{exp}\left(-\frac{\hbar \omega }{2~ k_BT}\right) \text{sinh}\left(\frac{\sqrt{2} \Omega  \hbar }{k_BT}\right); \nonumber \\[5pt]
 \rho_{55}=\text{exp}\left(-\frac{\hbar \omega }{2 k_BT}\right) \left(1+2 \text{Cosh}\left(\frac{\sqrt{2} \Omega  \hbar }{k_BT}\right)\right);~ \rho_{64}=-2 \sqrt{2}~ \text{exp}\left(\frac{\hbar \omega }{2~ k_BT}\right) \text{sinh}\left(\frac{\sqrt{2} \Omega  \hbar }{k_BT}\right); \nonumber \\[5pt]
 \rho_{66}=4~ \text{exp}\left(\frac{\hbar \omega }{2~ k_BT}\right) \text{cosh}\left(\frac{\sqrt{2} \Omega  \hbar }{k_BT}\right);~ \rho_{67}=-2 \sqrt{2}~ \text{exp}\left(\frac{\hbar \omega }{2~ k_BT}\right) \text{sinh}\left(\frac{\sqrt{2} \Omega  \hbar }{k_BT}\right); \nonumber \\[5pt]
 \rho_{74}=\text{exp}\left(\frac{\hbar \omega }{2~ k_BT}\right) \left(-1+2 \text{cosh}\left(\frac{\sqrt{2} \Omega  \hbar }{k_BT}\right)\right);~ \rho_{76}-2 \sqrt{2} \text{exp}\left(\frac{\hbar \omega }{2~ k_BT}\right) \text{sinh}\left(\frac{\sqrt{2} \Omega  \hbar }{k_BT}\right); \nonumber \\[5pt]
 \rho_{77}=\text{exp}\left(\frac{\hbar \omega }{2~ k_BT}\right) \left(1+2 \text{cosh}\left(\frac{\sqrt{2} \Omega  \hbar }{k_BT}\right)\right);~ \rho_{88}=\text{exp}\left(\frac{3 \hbar \omega }{2 k_BT}\right). 
\end{align}

From the above description, one observes that the system at high temperature is perfectly separable.  However, for intermediate temperatures, the system is in a mixed state and we have investigated the intensity pattern and photon-photon correlation of such a system. The positive frequency component of the electric field operator \cite{gsa, r17} is given by,
\begin{equation}
\label{eq:E}
\hat{E}^{(+)}=-\frac{e^{ikr}}{r}\sum_{j}\vec{n}\times(\vec{n}\times\vec{p}_{ge})e^{-i\phi_{j}}\hat{S}^{-}_{j},
\end{equation}
where $r=|\vec{r}|$, with $\vec{r}$ indicating the detector position, the unit vector $ \vec{n}=\frac{\vec{r}}{r} $ and $ \vec{p}_{ge} $, the dipole moment of the atomic  transition $ |e \rangle \rightarrow |g \rangle $. Here  $ \phi_{j} $ is the relative optical phase accumulated by a photon emitted at $ \vec{R}_{j} $ and detected at $ \vec{r} $. \newline

We also assume $ \vec{p}_{ge} $ to be oriented along the y-direction and $ \vec{n} $ in the x-z plane, resulting in vanishing $\vec{p}_{ge}.\vec{n}$. These assumptions, together with the  normalization,  give rise to dimensionless expressions for the amplitude as,

\begin{equation}\label{eq:E}
\hat{E}^{(+)}=\sum_{j}e^{-i\phi_{j}}\hat{S}^{-}_{j},
\end{equation}
resulting  in the following expression for the radiated intensity at $ \vec{r} $:

 \begin{flalign*}
I(\vec{r})=\left<\hat{E}^{(-)}\hat{E}^{(+)}\right>
=\sum_{i,j}\langle \hat{S}^{+}_{i}\hat{S}^{-}_{j}\rangle e^{i(\phi_{i}-\phi_{j})}, 
\end{flalign*}

\begin{equation}
=\sum_{i}\langle \hat{S}^{+}_{i}\hat{S}^{-}_{i} \rangle +
\left(\sum_{i\ne j}\langle \hat{S}^{+}_{i}\rangle \langle \hat{S}^{-}_{j}\rangle +  \sum_{i\ne j}(\langle \hat{S}^{+}_{i}\hat{S}^{-}_{j}\rangle - 
\langle \hat{S}^{+}_{i}\rangle \langle \hat{S}^{-}_{j}\rangle)\right )e^{i(\phi_{i}-\phi_{j})}.
\label{eqn:Int form}
\end{equation}
Thus, the characteristics of the intensity would depend on the incoherent terms $ \langle \hat{S}^{+}_{i}\hat{S}^{-}_{i}\rangle $, the non-vanishing of the dipole moments $\langle \hat{S}^{+}_{i}\rangle$, and the QCs of the form $\langle \hat{S}^{+}_{i}\hat{S}^{-}_{j}\rangle - \langle \hat{S}^{+}_{i}\rangle \langle \hat{S}^{-}_{j}\rangle $. \newline

We now take into account the thermal effects, where,  at finite temperature, the expectation value of an observable $\left<\hat{A}\right>$ takes the form

\begin{equation}
    \left<\hat{A}\right>=\text{Tr}\left(\hat{\rho}\hat{A}\right).
    \label{exp_value}
\end{equation}

In the line configuration, a system of three identical dipole coupled two-level atoms are placed symmetrically along a line with equal spacing $ d $ between adjacent atoms. For this topology,  $ \phi_{j} $,  the relative optical phase accumulated by a photon emitted at $ \vec{R}_{j} $ and detected at $ \vec{r} $ is
\begin{equation}
\phi_{j}(\vec{r})\equiv \phi_{j}=k\vec{n}.\vec{R}_{j}=jkd\sin{\theta}.
\label{eqn:phase line}
\end{equation}
The exact expression for the intensity for three atoms arranged along a line is derived  by combining Eqs. (\ref{eqn:Int form}) to (\ref{eqn:phase line}) and is given by

\begin{multline}
I=\langle E^{-}E^{+}\rangle=A~(B+C+D),\\[6pt]
\text{with}~~~~~~~~~~~~~~~~~~~~~~~~~~~~~~~~~~~~~~~~~~~~~~~~~~~~~~~~~~~~~~~~~~~~~~~~~~~~~~~~~~~~~~~~~~~~~~~~~~~~~~~~~~~~~~~~~~~~~~~~~~~~~~~~\\[4pt]
A=\frac{\text{exp}\left(-\frac{\hbar \omega }{2 k_BT}\right) \text{sech}\left(\frac{\hbar \omega }{2K T}\right)}{2 \left(1+2~\text{cosh}\left(\frac{\hbar \omega }{k_BT}\right)+8~ \text{cosh}\left(\frac{\sqrt{2} \hbar \Omega }{k_BT}\right)\right)},~~~~ B=3~\text{exp}\left(\frac{2 \hbar \omega }{k_BT}\right)-2~\text{exp}\left(\frac{\hbar \omega }{k_BT}\right) (-2+\text{cos}(2 \text{kd} \text{sin}(\theta))),~~~~~~~~~~~~~~~~~~~~~~~~~~~~~~~~~~~~~~~~~~~~~~~~~~~~~~~~~~~~~~~~~~~~~~~~~~~~~~~~\\[4pt]
 C=4 \left(2+\text{cos}(2 \text{kd} \text{sin}(\theta)+\text{exp}\left(\frac{\hbar \omega }{k_BT}\right) (4+\text{cos}(2 \text{kd} \text{sin}(\theta ))))\right) \text{cosh}\left(\frac{\sqrt{2} \hbar \Omega }{k_BT}\right),~~\text{and}~~~~~~~~~~~~~~~~~~~~~~~~~~~~~~~~~~~~~~~~~~~~~~~~~~~~~~~~~~~~~~~~~~~~~~~~~~~~~~~~~~~~~~~~~\\[4pt]
 D=4 ~\text{sin}^2(\text{kd} \text{sin}(\theta ))-8 \sqrt{2} \left(1+\text{exp}\left(\frac{\hbar \omega }{k_BT}\right)\right) \text{cos}(\text{kd} \text{sin}(\theta))~ \text{sinh}\left(\frac{\sqrt{2} \hbar \Omega }{k_BT}\right). ~~~~~~~~~~~~~~~~~~~~~~~~~~~~~~~~~~~~~~~~~~~~~~~~~~~~~~~~~~~~~~~~
 \label{Int_form}
\end{multline}

%%%%||||||||||||||||||||||||||||||||||||||||||||||||||||||||||||||||||||||||
\begin{figure}[H]
\centering
\includegraphics[width=18.0 cm,height=9.50 cm]{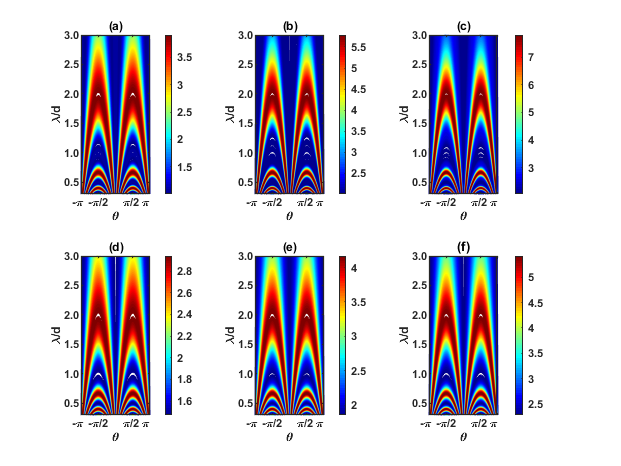} 
\caption{The variation of intensity is shown as a function of $\theta$ and $\frac{\lambda}{d}$ at $\frac{\omega}{\Omega}=1$ for (a) $N=3$, $T_1$, (b) $N=4$, $T_1$, (c) $N=5$, $T_1$, (d) $N=3$, $T_2$, (e) $N=4$, $T_2$, and (f) $N=5$, $T_2$, with $T_1=5\times10^{-3}\frac{\hbar\Omega}{k_B}$ and $T_1=\frac{\hbar\Omega}{k_B}$, and $N$ is number of atoms.}
\label{fig:Line theta lambda}
\end{figure}
%%%%||||||||||||||||||||||||||||||||||||||||||||||||||||||||||||||||||||||||

The analytical expression of photon-photon or intensity-intensity correlation is given by

\begin{equation}
g^{(2)}(0)=\dfrac{\langle E^{-}E^{-}E^{+}E^{+}\rangle}{\langle E^{-}E^{+}\rangle \langle E^{-}E^{+}\rangle}=\dfrac{\langle E^{-}E^{-}E^{+}E^{+}\rangle}{\langle E^{-}E^{+}\rangle^2}.
\label{eqn:g20}
\end{equation}
The numerator in Eq. (\ref{eqn:g20}) is given by
\begin{multline}
   \langle E^{-}E^{-}E^{+}E^{+}\rangle=N_1~(N_2+N_3),~~~~~~~~~~~~~~~~~~~~~~~~~~~~~~~~~~\\[4pt]
    \text{with}~~~~~~~~~~~~~~~~~~~~~~~~~~~~~~~~~~~~~~~~~~~~~~~~~~~~~~~~~~~~~~~~~~~~~~~~~~~~~~~~~~~~~~~~~~~~~~~~~~~~~~~~~~~~~~~~~~~~~~~~~~~~~~~~\\[4pt]
    N_1=\frac{\text{exp}\left(2~kd \text{sin}(\theta)+\frac{\hbar \omega - \sqrt{2} \hbar\Omega}{2k_BT}\right) \text{sech}\left(\frac{\hbar \omega }{2k_BT}\right)}{2 \left(1+2~\text{cosh}\left(\frac{\hbar \omega }{k_BT}\right)+8~ \text{cosh}\left(\frac{\sqrt{2} \hbar \Omega }{k_BT}\right)\right)},~~~~~~~~~~~~~~~~~~~~~~~~~~~~~~~~~~~~~~~~~~~~~~~~~~~~~~~~~~~~~~~~~~~~~~~~~~~~~~~~~~~~~~~~~~~~~~~~~~~~~~~~~~~~~~~~~~~~~~~~~~~~~~~~~~~~\\[4pt]
    N_2=-4 \sqrt{2} \left(-1+\text{exp}\left(\frac{2 \sqrt{2} \hbar \Omega }{k_BT}\right)\right) \text{cos}\left(\text{kd} \text{sin}(\theta)\right)+2 ~(2+\text{cos}(2~\text{kd} \text{sin}(\theta))),~~~~\text{and}~~~~~~~~~~~~~~~~~~~~~~~~~~~~~~~~~~~~~~~~~~~~~~~~~~~~~~~~~~~~~~~~~~~~~~~~~~~~~~~~~~~~~\\[4pt]
    N_3=\text{exp}\left(\frac{2 \sqrt{2} \hbar \Omega }{k_BT}\right) \left(4+3~ \text{exp}\left(\frac{2 \hbar \omega }{k_BT}\right)+2~ \text{cos}(2~\text{kd} \text{sin}(\theta))\right)+4~ \text{exp}\left(\frac{\sqrt{2} \hbar \Omega }{k_BT}\right) \text{sin}^2(\text{kd} \text{sin}(\theta )).~~~~~~~~~~~~~~~~~~~~~~~~~~~~~~~~~~~~~~~~~~~~~~~~~~~~~~~~~~~~~~~~~~~~~~~~~~~~~~~~~~~~~~~~~~~~~~~~~~
\end{multline}
Therefore,

\begin{equation}
g^{(2)}(0)=\frac{N_1~(N_2+N_3)}{I^2}.
\label{eqn:g20}
\end{equation}

%%%%|||||||||||||||||||||||||||||||||||||||||||||||||||||||||||||||||||||||||||
\begin{figure}[H]
\centering

\includegraphics[width=13.0 cm,height=9.0 cm]{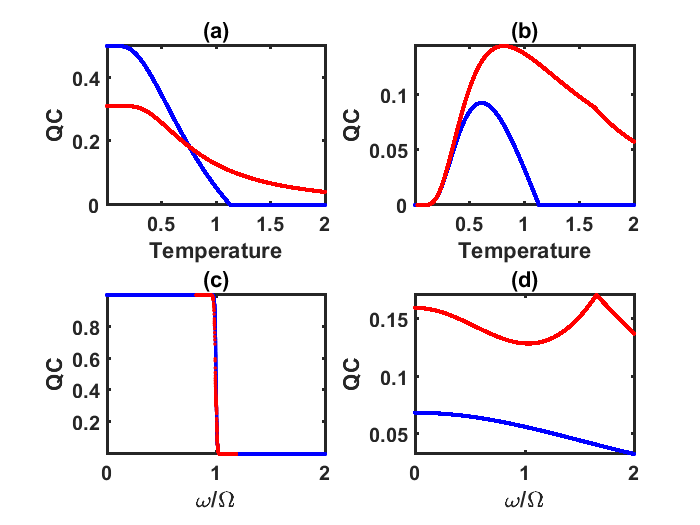}

\caption[]{(Color online) (a) Panels $a$ and $b$ show the variation of QCs (Concurrence (blue) and Discord (red)) as a function of temperature ($\times \frac{\hbar\Omega}{k_B}$) for panel $a$ $\frac{\omega}{\Omega}=1$ and panel $b$ $\frac{\omega}{\Omega}=2$. Panels $c$ and $d$ show the variation of QCs as function of $\frac{\omega}{\Omega}$ for panel $c$ $k_BT=5\times10^{-3}\hbar\Omega$ and panel $d$ $k_BT=\hbar\Omega$.}
\label{fig:two}
\end{figure}
%%%%|||||||||||||||||||||||||||||||||||||||||||||||||||||||||||||||||||||||||||

Fig. \ref{fig:two} shows the quantum correlations present in coupled two two-level atomic system. The behavior of entanglement and discord are same as coupled three atomic system, differing only in numerical value.

%%%%|||||||||||||||||||||||||||||||||||||||||||||||||||||||||||||||||||||||||||
\begin{figure}[H]
\centering

\includegraphics[width=13.0 cm,height=9.0 cm]{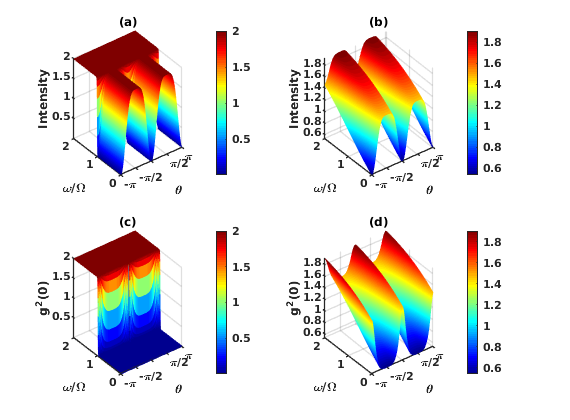}

\caption[]{Panel $a$ and $b$ investigate intensity pattern for photons emitted from two two-level atomic system as a function of $\frac{\omega}{\Omega}$ and $\theta$ for (a) $k_BT=5\times10^{-3}\hbar\Omega$ and (b) $k_BT=\hbar\Omega$. Panel $c$ and $d$ depicts photon-photon correlation $(g^2(0))$  as a function of $\frac{\omega}{\Omega}$ and $\theta$ for (a) $k_BT=5\times10^{-3}\hbar\Omega$ and (b) $k_BT=\hbar\Omega$., clearly showing vanishing of $(g^2(0))$ for maximally entangled sources. The value of $\frac{\lambda}{d}=2$ for all the plots.}
\label{fig:three}
\end{figure}
%%%%|||||||||||||||||||||||||||||||||||||||||||||||||||||||||||||||||||||||||||

Fig. \ref{fig:three} depicts the intensity and photon-photon correlation $(g^2(0))$ of coupled two two-level atomic system. It is clear from Fig. \ref{fig:two}(c) and Fig. \ref{fig:three}(c) that $g^2(0)=0$ for maximally entangled sources which means emitted photons are uncorrelated from each other, showing the complete anti-bunching effect.

\end{document}